\documentclass[apj]{emulateapj}
\journalinfo{\textsc{The Astrophysical Journal}}
\slugcomment{Received 2017 April 10; Accepted 2017 July 13}

\usepackage{url}
\usepackage{natbib}
\usepackage{booktabs}
\usepackage{threeparttable}
\usepackage{graphicx}

\usepackage[colorlinks,
        hypertexnames=false,
        pagebackref=false,
        linkcolor=blue,    
    citecolor=blue,        
    filecolor=magenta,     
    urlcolor=blue          
]{hyperref}                
\usepackage{amsmath}

\shorttitle{Successive Homologous CMEs from AR 12371}
\shortauthors{Vemareddy}

\begin{document}
\title{Successive Homologous Coronal Mass Ejections driven by shearing and converging motions in solar active region NOAA 12371}
\author{P.~Vemareddy}
\affil{Indian Institute of Astrophysics, II Block, Koramangala, Bengalure-560034, India}
\email{vemareddy@iiap.res.in}
\begin{abstract}
We study magnetic field evolution in AR12371 relating its successive eruptive nature. During the disk transit of seven days, the AR launched four sequential fast coronal mass ejections (CMEs) associated with long duration M-class flares. Morphological study delineates a pre-eruptive coronal sigmoid structure above the polarity inversion line (PIL) similar to Moore et al study. Velocity field derived from tracked magnetograms indicates persistent shear and converging motions of polarity regions about the PIL. While these shear motions continue, the crossed arms of two sigmoid elbows are being brought to interaction by converging motions at the middle of PIL, initiating tether-cutting reconnection of field lines and the onset of CME explosion. The successive CMEs are explained by a cyclic process of magnetic energy storage and release terming ``sigmoid-to-arcade-to-sigmoid" transformation driven by photospheric flux motions. Further, the continued shear motions inject helicity flux of dominant negative sign, which contributes to core field twist and its energy by building a twisted flux rope (FR). After a limiting value, the excess coronal helicity is expelled by bodily ejection of the FR initiated by some instability as realized by intermittent CMEs. This AR is in contrast to the confined AR12192 with predominant negative signed larger helicity flux but very weaker (-0.02turns) normalised coronal helicity content. While predominant signed helicity flux is a requirement for CME eruption, our study suggests the magnetic flux normalized helicity flux as a necessary condition accommodating the role of background flux and appeals a further study of large sample of ARs.
\end{abstract}

\section{Introduction}
Triggering of coronal mass ejections (CMEs) have mostly concentrated on the problem of evolution of a magnetic field in very tenuous highly conducting plasma of solar corona \citep{forbes1984, linj2000, linj2003, forbes2006}. This evolution is driven by slow (compared to the Alfven velocity) motions of the footpoints at the photosphere. When an active region (AR) emerges, line-of-sight motions in the early phase and horizontal (includes shear and/or converging and/or proper) motions after rapid emergence phase dominate \citep{liuy2012,vemareddy2015a}. To model the eruptive scenario of ARs under a particular evolving condition of boundary motion, numerical models have been constructed, vis., emerging conditions of motion \citep{fany2003,fany2004,gibson2006,piyali2013,archontis2014}, shear dominated motions \citep{antiochos1999,amari2003a} and converging motions \citep{amari2003b,amari2010}. All of these models are based on observationally \citep[e.g.,][]{tanaka1973, machado1986, priest2002, linj2003,tian2008b} valid physical concept that the footpoint motions predominantly contribute to coronal helicity budget to form a twisted flux rope (FR) during or before its ejection as CME. Despite this significant progress in the past decade combining observational and numerical modeling/simulation efforts, the relation between the energy build-up process by the flux motions and the initiation of the eruption is still an outstanding question.   

As soon as the magnetic system of the AR has reached a state of sufficiently adequate energy by these flux motions, its sudden release requires suitable triggering mechanism, which has gained significant interest in the past decade. Several ideas have been framed for explaining the onset mechanism of the eruption \citep{klimchuk2001, forbes2006, moore2006}. Internal tether-cutting reconnection in a sheared core field of single bipolar region \citep{moore1980,moore2001}, external tether cutting or breakout reconnection of overlying field with the low lying core field in a quadrapolar region \citep{antiochos1999}, flux cancellation \citep{martin1985, martin1989,martens2001}, emergence of twisted FR from below the surface \citep{leka1996}, and ideal MHD kink/torus instability \citep{torok2004,kliem2006} are few important mechanisms proposed to explain observed eruption process. Due to a wide variety of dynamic processes in complex ARs, it is difficult to identify a particular triggering mechanism responsible for a given eruption, however. 

Some ARs show rapid succession of CMEs and flares over a time scale of minutes to hours  \citep{gopalswamy2005}. As the time scale is too small compared to typical time scale for energy build up in ARs, the rapid succession of flares and CMEs therefore represents a fragmented energy release. The AR 9236 was reported to produce recurrent CMEs at an average time period of 10 hour \citep{gopalswamy2005} and the associated flares were not long decay events (LDEs). In such cases, studies also speculate role of preceding CME in creating the conditions for a subsequent event \citep{gopalswamy2004}. However, study of the same AR by \citet{nitta2001} suggested the emerging magnetic flux being responsible for repeated CMEs. In contrast, successive CMEs also occur from source ARs in a time scale comparable to energy build up by footpoint motions. A best example was AR 8038 producing recurrent CMEs in few days apart. In this decaying AR, studies report that prolonged flux cancellation by converging motions and subsequent magnetic gradient increase about the polarity inversion line (PIL) introduced energy build-up in the AR magnetic system that being erupted to CMEs/flares \citep{shibu2000, liy2004, liy2010}. Another recent observational report claims the formation of four successive homologous flux ropes during the evolution of AR 11745 but only last event becomes CME \citep{LiZhang2013}. Further recent reports in AR 11158 suggest that the shear and rotational motions of the observed fluxes played significant role in transient activity with flares and CMEs \citep{vemareddy2012b,sunx2012}. A similar study of AR 12158 concludes that the two successive CME eruptions being triggered by helical kink-instability under the driving conditions of predominant sunspot rotation in a time scale of days \citep{vemareddy2016b}. 

ARs with repeated CMEs are not common and even exist on the disk, simultaneous and uninterrupted photospheric and coronal observations may not available. Because of this fact not many observational investigations of successive eruptions from source AR appear in the literature. Therefore a general understanding on the relation between nature of surface flux motions and a particular triggering mechanism of repeated CMEs remains still elusive. Such studies deemed further understanding of the connection between the magnetic field evolution and the most favoured triggering mechanism for solar eruptions. This is the prime motivation of the present work on AR 12371 launching successive CMEs on side of the solar disk.  Further, occurrence of homologous CMEs generally refers to existence of flux rope and its eruption \citep{LiZhang2013,vourlidas2013}. Therefore, studying the AR magnetic evolution will throw some light on the flux rope formation mechanism, in particular during or before eruption that in turn decides the nature of triggering mechanism whether resistive reconnection related or ideal magnetohydrodynamic (MHD) instability based. These aspects of eruption process have great importance from the point of space weather prediction. Investigation of several such AR cases are also useful to reveal insightful information separating from their non-erupting counterparts \citep{sunx2015}. In Section~\ref{sec2}, we describe an overview of CME observations, and their initiation mechanism in Section~\ref{sec3}. The associated magnetic evolution is studied in Section~\ref{sec4} and compared with the one in AR 12192 in Section~\ref{comp}. We conclude the study with a discussion on the role of energy/helicity storage scenario relating the eruptive nature in  Section~\ref{disc}.   
\begin{table*}[!ht]
	\centering
	\begin{threeparttable}
		\caption{Major CME events from AR 12371 during the disk transit}
		\begin{tabular}{ccccc}
			\hline 
			event  & Time\tnote{a} &initiation time\tnote{b}  & LASCO speed (km/s)\tnote{c} & associated flare (time)\tnote{d} \\ 
			\hline 
			CME1&2015/06/18 17:24& 15:05  & 1305  & M3.1 (16:25)  \\ 
			CME2&2015/06/21 02:36& 00:45 & 1366 & M2.2 M2.7 (01:02, 02:00)  \\ 
			CME3&2015/06/22 18:36& 16:15 &  1209& M6.5 (17:39) \\ 
			CME4&2015/06/25 08:00& 07:30  & 1627 & M7.9 (08:02)  \\ 
			CME5&2015/06/24 14:24 & 14:30 & 399  & C5.6 (15:12) \\ 
			CME6&2015/07/01 14:00 & -- & 1435 &  away from visible disk\\ 
			\hline 
		\end{tabular} 
		\begin{tablenotes}
			\footnotesize
			\item[a] first lasco C2 appearance
			\item[b] begining time of EUV brightening
			\item[c] Obtained from \url{http://cdaw.gsfc.nasa.gov/CME_list/UNIVERSAL/2015_06/univ2015_06.html}
			\item[d] begin of GOES X-ray flux
		\end{tablenotes}
		\label{tab1}
	\end{threeparttable}
\end{table*}
\begin{figure*}[!ht]
	\centering
	\includegraphics[width=.99\textwidth,clip=]{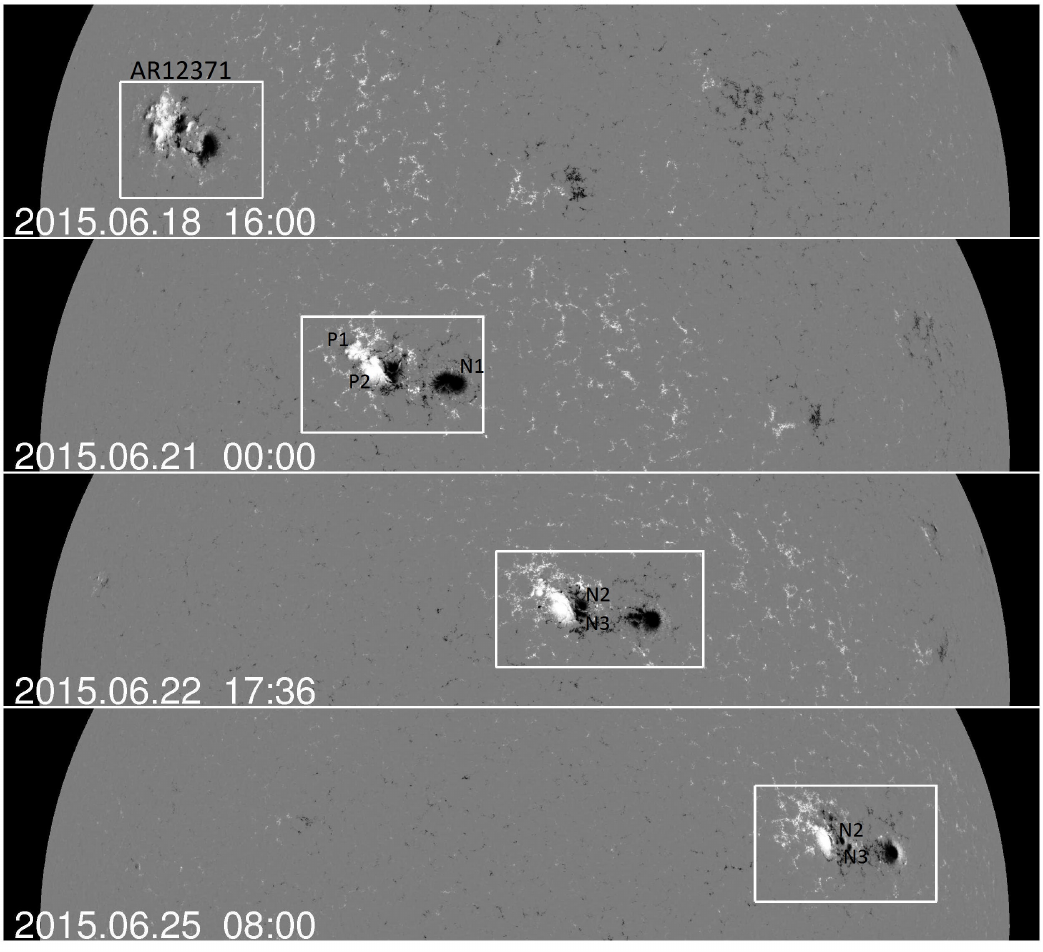}
	\caption{Position of the AR 12371 (rectangular inset in respective panels) on the solar disk during the major CME eruptions on four different days. Note shearing and coverging motion between the following polarities of opposite sign (P1, P2 and N2, N3), while they disintegrate and diffuse in area over time. }
	\label{fig1}
\end{figure*}

\section{Overview of CMEs from AR 12371}
\label{sec2}
The major source of observational data used in this study is Solar Dynamics Observatory \citep{pesnell2012}, which provides full magnetic field measurements and multitude of coronal observations uninterruptedly. The region of interest is recurrently CME producing AR 12371, which was a pre-emerged region first appeared on the disk at $12^\circ$N on June 16. In Figure~\ref{fig1}, we show the longitudinal extent of line-of-sight (LOS) magnetic field of the AR, obtained from Helioseismic Magnetic Imager (HMI; \citealt{schou2012}) on four different days of four major eruptions. For the convenience of description, we label the polarity regions according to sign. From these LOS magnetic field observations, the AR’s successive passage reveals presence of leading negative flux (N1) with a following negative (N2, N3) and positive (P1, P2) flux regions. The inner bipole (N2, N3) is seen with large shear and converging motion with respect to (P1, P2), and its flux distribution becomes diffused and disintegrated in successive days, while the leader polarity is increasingly seperated from the following polarity. Especially, the following negative polarity appears with a light bridge and splits into N2 and N3 while being in shear motion towards south direction.

\begin{figure*}[!ht]
	\centering
	\includegraphics[width=.85\textwidth,clip=]{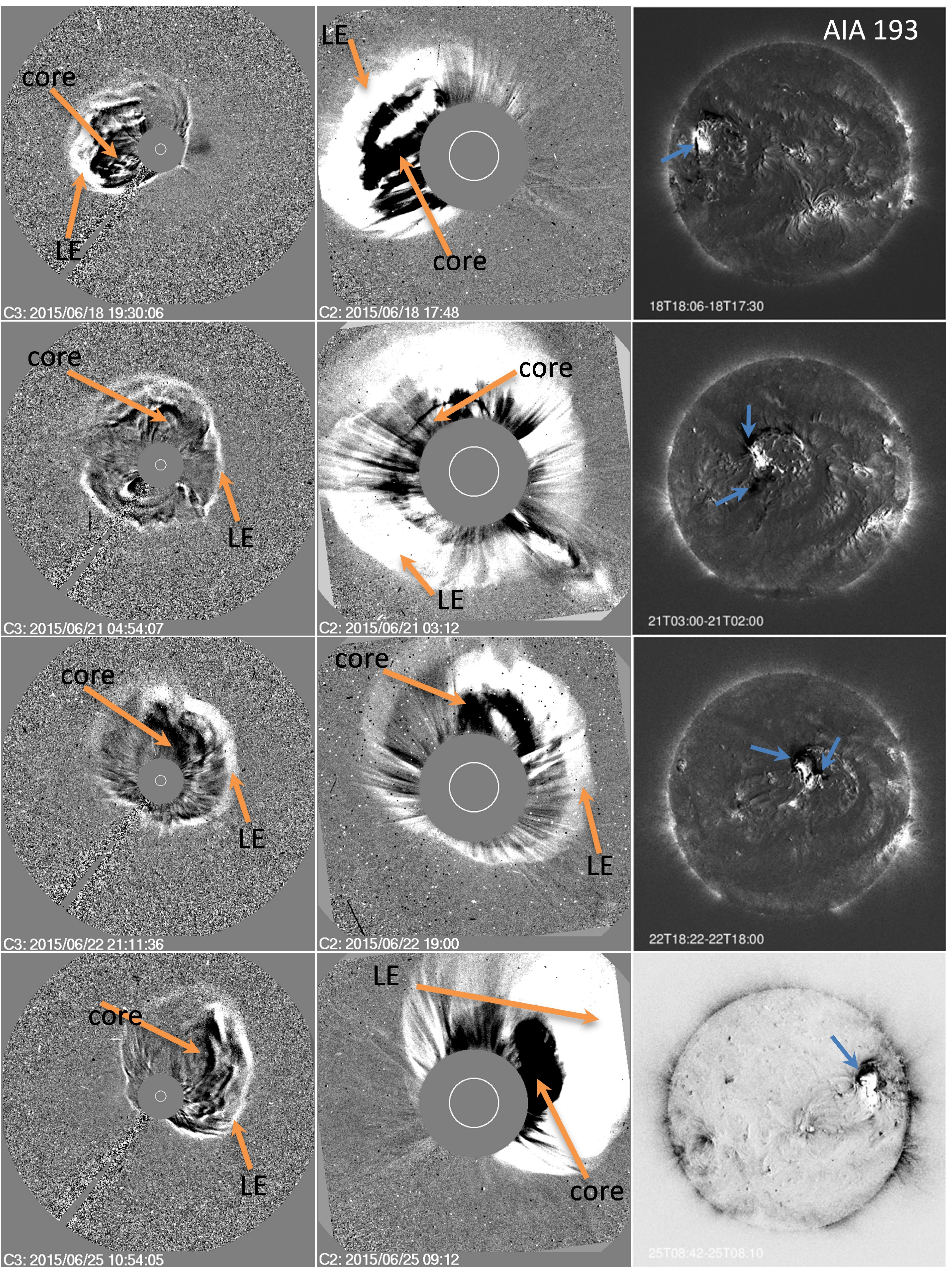}
	\caption{Time diffence images in LASCO/C3 (1st column), C2(2nd column), AIA 193 (3rd column) of four major CMEs (in each row) occurred in AR 12371. Leading edge and core parts are marked by arrows, whereas arrows in last column point dimming regions at the legs of the sigmoid. Note that CME1 and 4 are deviated from sun-earth line while CMEs 2 and 3 are full halo corresponding to the AR position on the Sun.}
	\label{fig2}
\end{figure*}

The whitelight CME observations are obtained from \textit{Large Angle and Spectrometric Coronagraph} (LASCO, \citealt{brueckner1995}) onboard the \textit{Solar Heliospheric Observatory} (SoHO). A summary of major eruptions took place in this AR and their observational signatures are given in Table~\ref{tab1}. The CME speed is measured in plane-of-sky and some of these quantities are taken from LASCO CME catalog, GOES flare catalog. The LASCO observations of these four CME cases (in each row) are depicted in Figure~\ref{fig2}.  CME1 occurred on June 18 when AR was at $48^\circ$ East longitude from the central meridian, and emerged on East limb at 17:24 UT in C2 and 18:30 UT in C3 field of view (FOV). The CME is catagorised as fast one from kinematics of leading edge traveling at a speed of 1305 km/s in the LASCO FOV. This CME is deflected from the sun-earth line since the AR is situated well away from the disk center. CME2 occured on June 21 when the AR is at $15^\circ$ East from the central meridian. It appears from 02:30 UT in C2 and 3:30 UT in C3 FOV. It is associated with successive M2.0 (01:02 UT), M2.7(02:04 UT) GOES class flares. The CME travels along the sun-earth line at a speed of 1366 km/s in LASCO FOV. Because the halo CME moves along LOS direction, this speed very likely represents the CME lateral expansion rate.

CME3 is intiated at 17:30 UT on June 22, when AR is at $7^\circ$ west of central meridian. It is associated with M6.5 flare occurred 17:30 UT and headed towards earth at a projected speed of 1209 km/s. It appears in C2 from 18:30 UT and in C3 from 19:18 UT onwards. CME4 is initated from 07:30 UT on June 25 when AR was at $42^\circ$ East from central meridian. It is accompanied by a major flare of M7.9 triggered at 08:02 UT. Since the AR was away from disk center, the measured travel speed, which is 1629 km/s, is subjected to small projection  effect. CME5 occured on June 24, which is relatively minor event, preceding a C5.6 flare and slow outward motion. CME6 occured on 1st July  when the AR was positioned away from the west limb on the other side of the disk, which is classified as fast having a speed of 1435 km/s. For reasons of having data with less projection effects, we analyzed first four events in the later sections. 
To show the extreme ultra-violet (EUV) dimming, we plot in the last column of Figure~\ref{fig2}, difference images of AIA 193\AA~observations taken immediately after CMEs launch from the AR. EUV dimming, especially symmetric double dimming, has attracted a lot attention interpreting that the two dimming regions are the footprints of an expanding FR still connectted to the Sun and experiencing significant density depletion \citep{thompson1998, webb2000, yurchyshyn2006b, liuchang2007, attrill2008}. This density depletion at the sigmoid/FR legs causes void of EUV emitting plasma and creates intensifying dark regions when FR takes off the Sun. As marked by arrows, the images clearly show these dark regions. CME2 show exact symmetric double dimming pattern as lobes of the upward lifting FR. However in the rest of the events, the dimming is on a side reflecting the effect of LOS integration of EUV emission that the AR being at different longitudes away from central meridian. Although all these dimming cases do not fit with exact symmetric double dimmings, the observations demonstrate the FR structure of the CME as we see FR formation during a self amplifying triggering mechanism of the CME.

\section{Initiation and Triggering Mechanism of eruptions }
\label{sec3}
In all the four CME cases, the initiation mechanism is similar as they are occuring recurrently from the same polarity inversion line (PIL) and magnetic field structure. Given two competent models of FR \citep{rust1996,roussev2003,torok2005} and sheared arcade \citep{antiochos1998,antiochos1999,amari2003a,amari2003b, amari2010}, the choice is more biased to the later because the preeruptive morphology in coronal observations is not fit with unambiguous FR topology as those generic examples considered in the literature \citep{green2007, gilbert2007,zhangj2012}.  In this context, a flux rope is defined as a collection of twisted magnetic field lines spiraling around the same axis by more than one full turn. The sheared arcade is an arcade of field lines that runs nearly parallel to the PIL rather than crossing right over the PIL. The FR models consider a preexisted twisted FR before eruption, and is initiated by an ideal MHD instability. A sheared arcade is the initial configuration in the other model, which is being build by motions parallel to PIL but opposite in direction at the footpoints. Both the sheared and twisted fields are stored energy configurations with non-zero magnetic helicity. Sigmoids, filaments are mostly considered as precursor features in such configurations \citep{pevtsov2003,canfield1999}. In this AR 12371, the preerupive configuration during all CME eruptions is a sigmoid as revealed by coronal morphology. Filament segments are also observed in 304\AA~images in some cases. A key feature of a filament channel is that it is a region of dominant horizontal field where the field on either side points in the same direction. As a result, filament channels are interpreted as locations of strong magnetic shear and highly non-potential magnetic fields \citep{mackay2010}. The observed sigmoid core is associated with inner bipole (P1, P2; N2 N3) (inner bipole, from here onward). As these polarities are seen with large shear motions (See Figure~\ref{fig1},\ref{fig8}) with their disappearing flux content, the build up of sheared arcade with a central core is in evitable \citep{antiochos1994, aulanier2002, antiochos1999, amari2003a} and is in accordance with the sheared arcade scenario.  

In addition to this sheared PIL, the orientation, separating motion of N1 from follower polarities produce further stress in the field lines of large scale covering end-to-end polarities, which we expect to build the non-potential energy in the entire magnetic structure of the AR. Moreover, a visual inspection of magnetograms gives an impression of converging motion of (N2, N3) towards (P1,P3) (see Figure~\ref{fig7}), which besides shear motions played prime role in building the sigmoid and subsquent eruption. With the above two important supporting points, we suggest the eruptions are driven by shearing and converging motion of oppositive polarities. Importantly, the continuous shear motions also transform the post eruption arcade (PFA) from the previous eruption further into sheared arcade to build energy for the next eruption. Flux convergence and cancellation have been shown to be important for filament channel formation according to the observations reported by \citet{martinsf1998}. \citet{WangMuglach2007} argue that the flux cancellation between opposite polarity elements removes the normal component of the field leaving the component parallel to the PIL which builds gradually to form the axial field of the filament channel. Similarly, \citet{amari2010} showed that flux cancellation may transform a sheared arcade into a stable or unstable FR. \citet{mackay2006} showed that flux cancellation at the PIL between bipoles may result in the formation of a FR, its elevation and consequence reconnection below it and subsequent ejection.
\begin{figure*}[!ht]
	\centering
	\includegraphics[width=.99\textwidth,clip=]{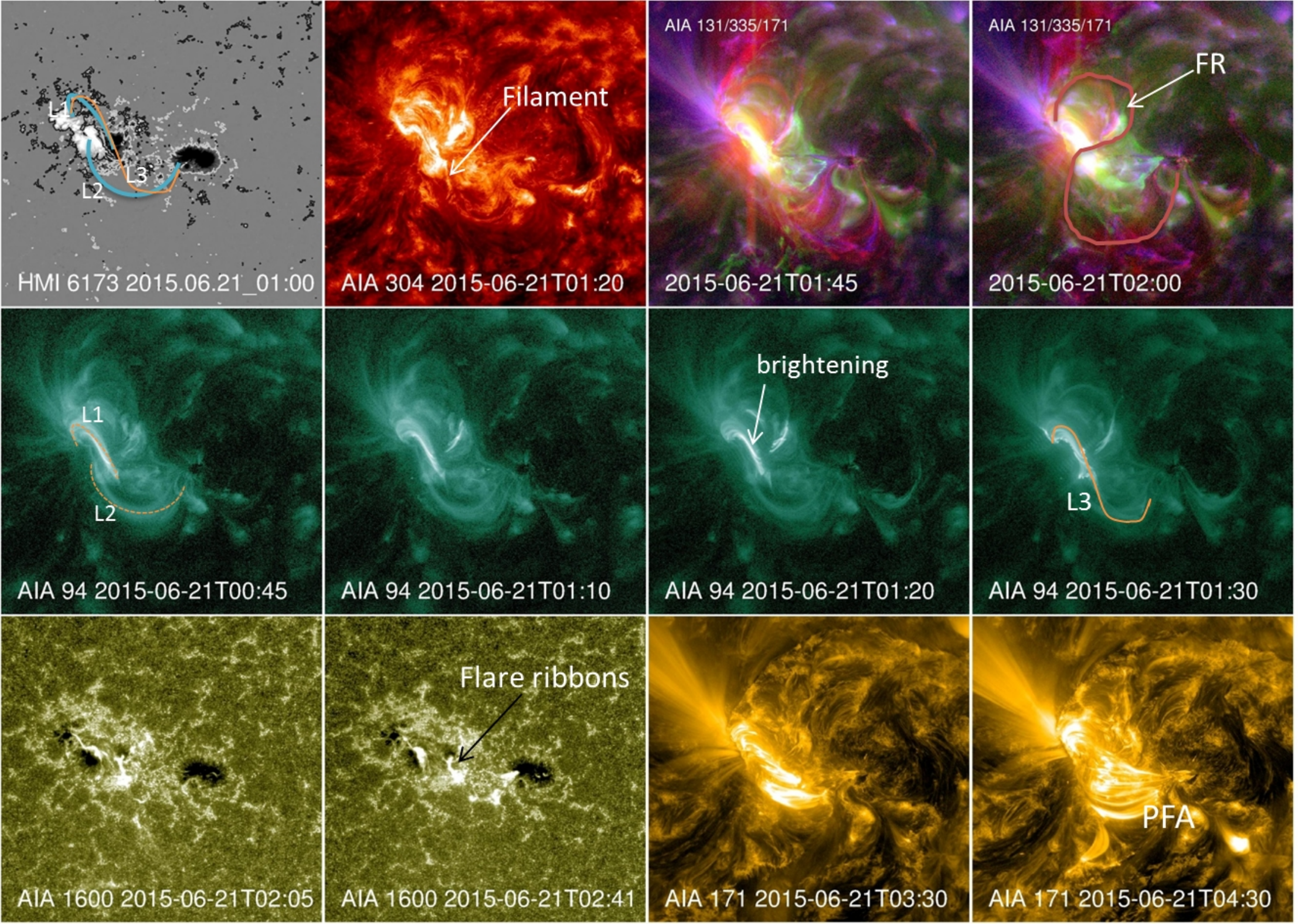}
	\caption{Observations during initiation of CME2. Top row: Co-aligned maps of HMI magnetic field and AIA 304, composite 131(red)/335(green)/193(blue)\AA~passbands. Trace of dark filemant is identifiable in AIA 304\AA~ image. Trace of reconnecting two loop sets  L1,  L2 (turquoise curves) and their product loop system L3 (FR, orange curve) are depicted on magnetogram. Due to direct LOS view, inverse-S FR is obvious in composite images.  Second row: Snopshots of AIA 94\AA~ observations showing slow phase tether-cutting reconnection of two inverse-J section loops (L1 and L2) and formation of FR (L3). Increased brightness persists with L1 (dotted orange curve) as reconnection progreses to form L3. Third row: AIA 1600\AA~ snapshots (upper photosphere) and coronal PFA in AIA 171 during posteruption phase. Flare ribbons in upper photosphere corresponds to foot prints of PFA and takes inverse-S shape morphology. FOV in each panel is $450\times480$\,arcsec$^2$.   }
	\label{fig4}
\end{figure*}

In slowly evolving corona, the eruption of the sigmoid is explained by initiation and trigger mechanisms. They occur in a time scale of less than an hour or so, far less than that of build up of the sigmoid. From similar studies of ARs \citep{green2009,green2011,vemareddy2015b}, the mechanism that best explains the eruptions in our AR 12371 is tether cutting model \citep{moore1980,moore2001,moore2006}. In this model, the pre-eruptive sigmoid consists of a central shear core about PIL and two oppositely curved loops as magnetic elbows. These two elbow arms shear past each other along the middle stretch of the PIL. At some (can be critical) stage, these oppositely curved elbows also appear as 
-sections, with a sharp interface between the elbow arms, also called crossed arms, rooted in opposite polarities about the middle section of the PIL. When these arms are pushed further against each other by converging motions, they begin to reconnect at the interface. The newly connected field lines exit from the reconnection site in the corona, one set escapes upward as twisted FR and the other is downward release field lines as short sheared loops low over PIL. This process of reconnection occurs explosively. In the following, we compare the coronal morphology to explain the details of this triggering model in the observed CME eruptions. 

\subsection{Evidence for tether-cutting reconnection}

\begin{figure}[!ht]
	\centering
	\includegraphics[width=0.49\textwidth]{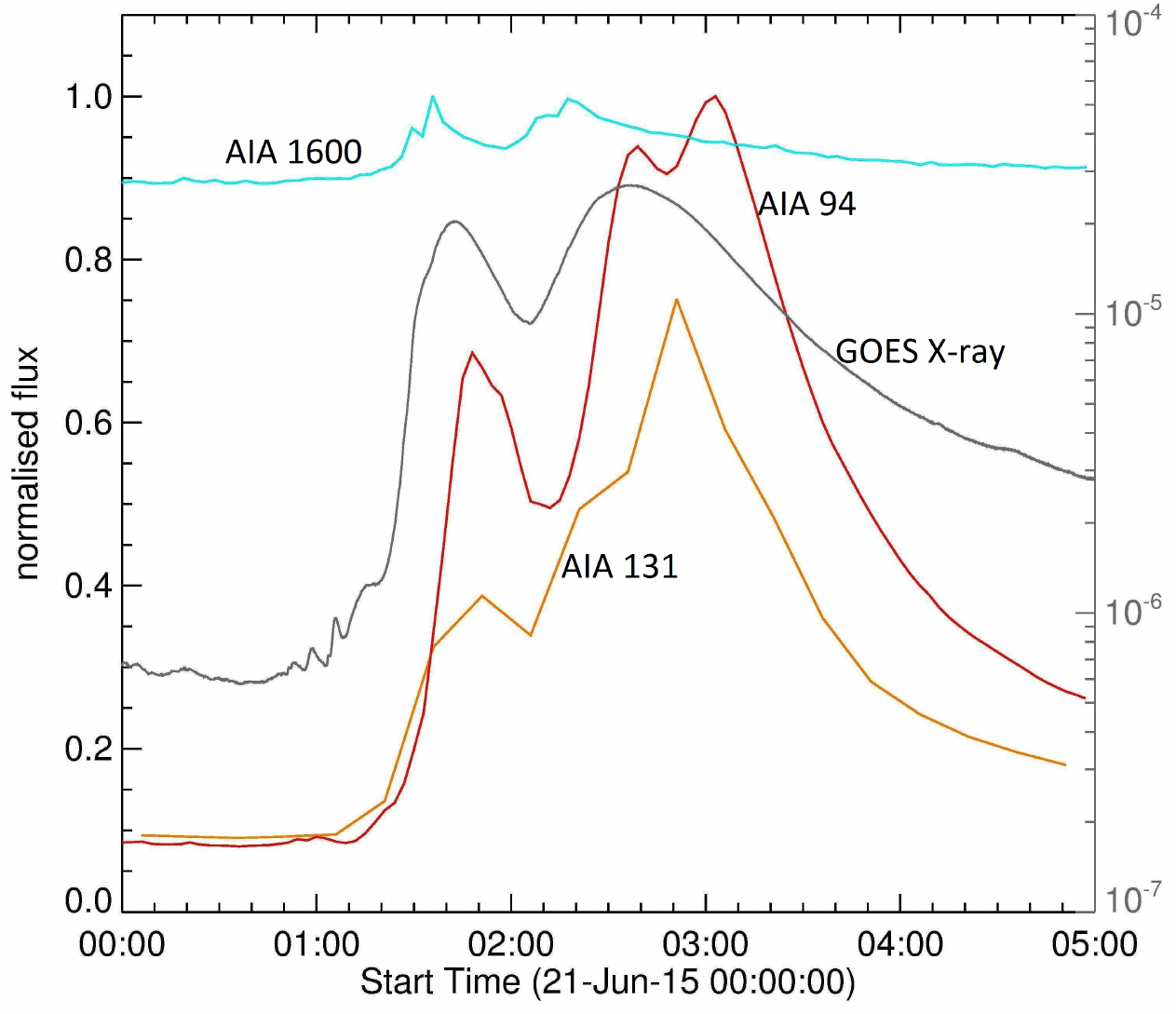}
	\caption{GOES X-ray flux and light curves of AR 12371 in AIA 94, 131 and 1600\AA~wave bands during the onset of CME2. Double peaks in these curves corresponds two phases of internal tether-cutting reconnection. Already formed FR in the first phase remains confined to coronal environment during decreasing period of these light curves, which then pushed up by the second phase  reconnection product and then subsquent eruption.}
	\label{lightcurve}
\end{figure}
In Figure~\ref{fig4}, we plot SDO multiwavelength observations during the onset of the CME2. The two instrument (AIA and HMI) observations at different times are co-aligned so as to map the footpoints of coronal loops in magnetic field observations. By the time of this eruption, the AR is located near disk center, so coronal structure and magnetic field distribution are less subjected to projection effects. Compared to previous event, positive polarity moved down to southward and the sigmoid with two inverse J-section loops now, with their crossed legs in the opposite polarity about the PIL, are more distinctly visible in AIA 94 snapshots (middle row panels). Note that we use AIA 94 \AA~observations instead 131 \AA~as in other events. AIA 304\AA~observations show contonuous filament trace (similar to FR) from N1 through PIL to P1 and is a signature of being supported by the crossings of L1 and L2. This case is similar to that studied by \citet{vemareddy2012a} who considered the filament channel as FR and interpreted role of tether-cutting reconnection in triggering its explosion. 

The AIA 94\AA~images reveal the initiation of eruption at 00:45 UT on June 21 with the brightening of loop set L1. From then, the brightening of L1 increasingly intensifies and the distiction between L1 and L2 disappears by the emergent of forming continuous loop set L3 (FR). It connects the far ends of L1 and L2 and is a signature of internal tether-cutting reconnection at the interface of crossed legs. The FR is seen clearly in composite images of AIA 131/334/193 as it further develops by run-away tether-cutting reconnection and expands in height subsequently.

The AIA as well as GOES flux show an interesting two phase evolution of light curves. We plot light curves of the AIA 94, 131, 1600, and GOES X-ray flux in Figure~\ref{lightcurve}. These light curves begin increasing in intensities starting from 00:50UT reaches a maximimum at 01:40UT and then they decrease slightly upto 02:00 UT from where they further increase. In the GOES scale of magnitude, they are recognised as M2.0 flare at 01:02UT and M2.7 flare at 02:04 UT. We carefully studied the imagery information in hot wavelengths and found that the observed light curves essentially imply FR height profile as soon as it formed. In the first phase, the forming FR L3 reaches certain height while producing reconnection related heating in thinning current sheet under it and remains confined to the AR magnetic environment in the decreasing period of the light curves. Trace of this FR is depicted with an inverse-S curve in AIA131/193/335 composite image panel. Its loss of confinement and full eruption occurs only in the second phase with the burst of second phase of internal tether-cutting reconnection (02:04 UT) producing second set of loops pushing the earlier FR and unleasing the explosion at 02:25 UT.

\begin{figure*}[!ht]
	\centering
	\includegraphics[width=.7\textwidth,clip=]{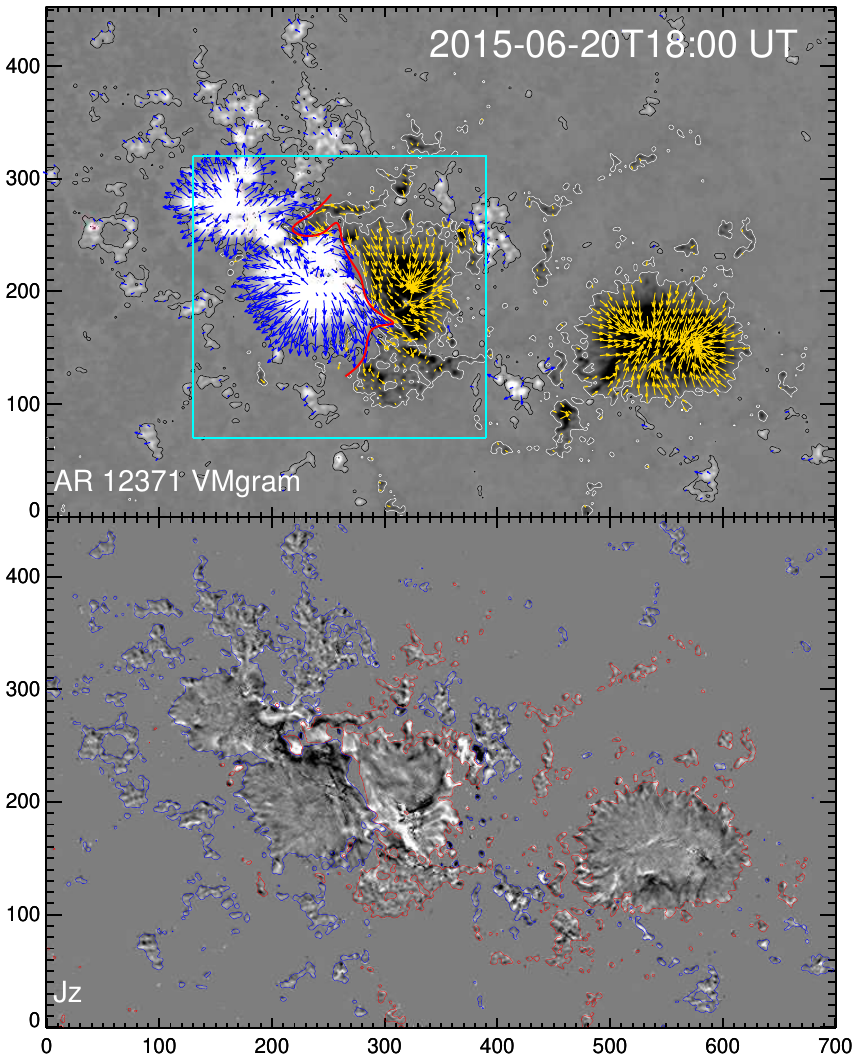}
	\caption{Top: Sample vectormagnetogram of AR 12371 on June 20 at 18:00UT. Background is $B_z$ map overplotted by transverse field vectors in blue (yellow) in regions bounded by contour of 100G (-100G). Region enclosed by cyan rectangle refers to inner bipolar region. Red curve is PIL. bottom: Vertical current ($J_z$) distribution scaled within $\pm30mAm^{-2}$. Axis units are in pixels of 0.5arcsec size. 
	} 
	\label{fig7}
\end{figure*}

As the eruption is taken place in two phases with the upward released FR lifting off, new bright low sheared loops appearing below. It is worth to mention that the filament trace after eruption disappeared from AIA 304\AA~images and the immediate post eruption arcade still remain sheared (last two panels in third row). A further relaxation of PFA continues for an extended period upto 5:00UT and becoming less sheared.  As seen in AIA 1600\AA~observations, flare ribbons begin appearing only after second phase of reconnection (01:50 UT onwards) and become intensively bright corresponding to peak (impulsive) phase of the flare at 02:30UT (third row panels). The ribbon morphology traces inverse-S shape although brightening is not continuous all along inverse-S path.  Unlike standard 2D flare model (CSHKP), this observed ribbon morphology, instead of two ribbons parallel to PIL, became a debate in several studies \citep{demoulin1996}. The issue leads to extension of 2D to 3D eruptive flare model,  linking the legs of erupting FR with the flare ribbons \citep{aulanier2012}. In this extended model, the shape of erupting FR and the sigmoid identifies cospatial ribbon morphology with hook shaped quasi separatrix layers. Starting from EUV sigmoid configuration, its transformation through initation, CME, flare phases is known as ``sigmoid to arcade" evolution.  

Similarly, we analyzed the EUV observations during the onset of CME1, CME3, CME4 for which the morphology of plasma loops well fit into tether-cutting model. The corresponding figures are referred at \url{ftp://ftp.iiap.res.in/vemareddy/movies/}.

\section{Evolution of magnetic non-potential parameters}
\label{sec4}
The non-potential nature of magnetic field is a measure of AR eruptivity and reveals the connection of flux motions with the energy build up in the AR magnetic structure. It is estimated by several parameters; viz, magnetic flux ($\Phi$), net vertical current ($I=\int J_z dS$), twist parameter $\alpha_{av}$, helicity, energy injection rate denoting $dH/dt$, $dE/dt$ respectively. For details of computing these parameters, we refer to \citet{vemareddy2015a, vemareddy2016b, vemareddy2017a} (also see citations therein). For this purpose, we used HMI vector magnetic field observations of AR patch. The vector magnetic field in the AR patch is deduced after a pipelined procedure of stokes vector inversion and ambiguity resolution \citep{borrero2011,hoeksema2014,bobra2014}. These disambiguated vector observations of the AR patch in the native coordinate system (latitude, longitude) are remapped to disk centre by cylindrical equal area (CEA) projection method such that the AR patch center matches the disk center. This is a spherical transformation accounting the foreshortening effect and the final image of the AR patch appears as if one is observing directly overhead \citep{calabretta2002}. The field vectors are then transformed to heliocentric spherical cooardinate system resulting ($B_r$, $B_\theta$, $B_\phi$) which are provided as \texttt{hmi.sharp\_cea\_720s} data product. These field components ( $B_r$, $B_\theta$, $B_\phi$) in the Heliocentric spherical coordinate system can be approximated as ($B_z$, -$B_y$, $B_x$) in the local heliographic Cartesian coordinate system (see appendix section in \citealt{sunx2013a}) when the field of view is small, which is acceptable for many studies in the Cartesian coordinates. Note that $B_z$ is essentially the radial component $B_r$ in the remapped series which we refer as vertical component of {\bf B}. To reduce inconsistencies due to measurement errors, we set a threshold limit of 150G for tranverse field components and 50 G for $B_z$ component. 

\begin{figure*}[!ht]
	\centering
	\includegraphics[width=.8\textwidth,clip=,]{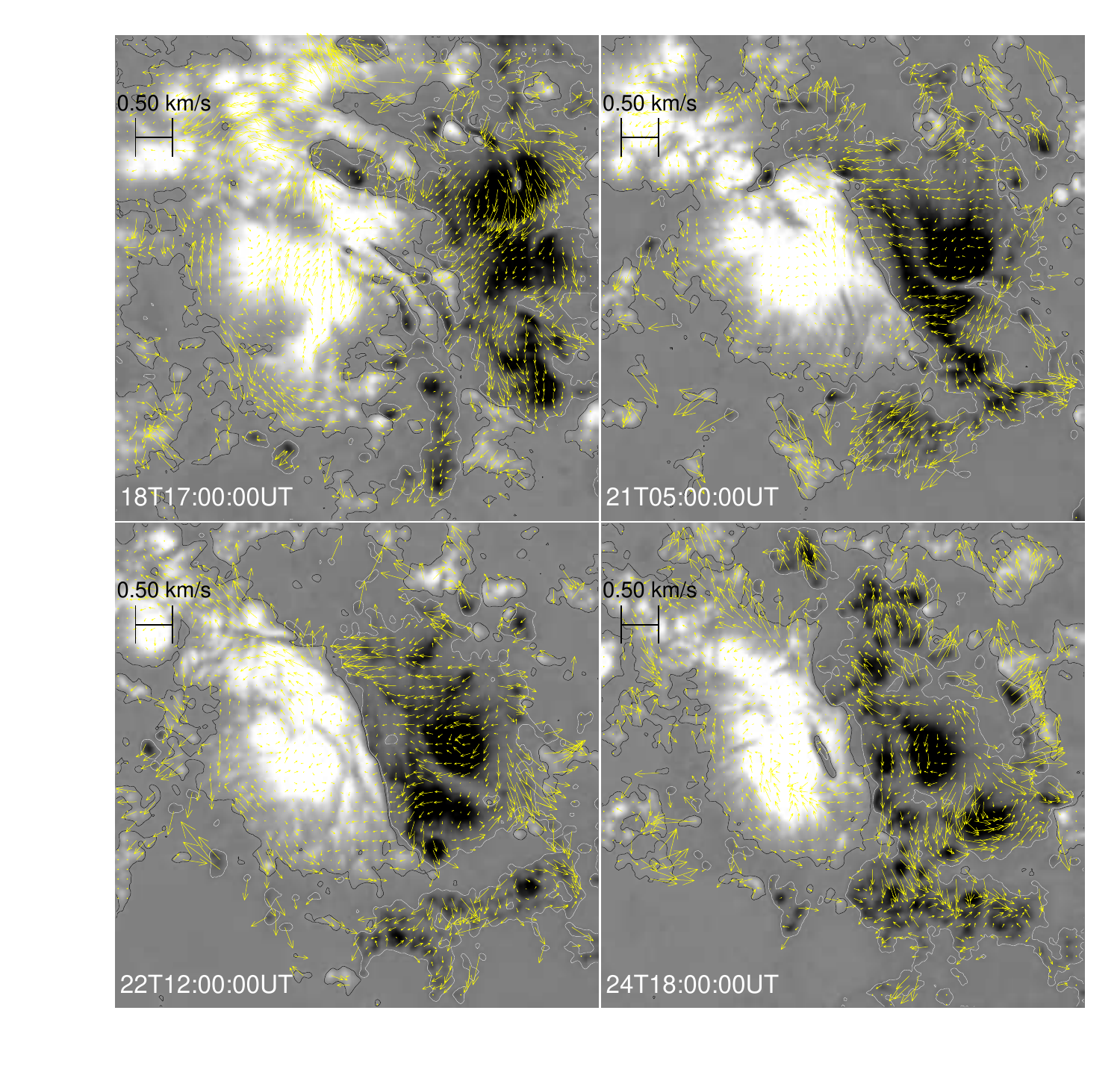}
	\caption{Horizontal velocity of flux motions derived from DAVE4VM technique at epochs in AR 12371. FOV covers only inner bipolar region indicated by rectangle in Figure~\ref{fig7}. In all panels, background image is $B_z$ component overlaid by contours at $\pm120$G and arrows of horizontal velocity field. Direction of arrows located in negative polarity regions (N2, N3) largely imply shearing and converging motion.} 
	\label{fig8}
\end{figure*}

Further, any projection method (here CEA) distort the spherical geometry at some level, basically the area, shape, direction, distance and scale. For a typical AR whose size is much smaller than the solar radius, the difference that results from applying different projection methods is small. \citet{liuyang2014} evaluated the impact of this difference that could bring in helicity flux calculation. The maximum difference they found is 0.7\% at an average of 0.36\%, thus concluding the small impact of projection method in our computations of different non-potential parameters. 

A typical vectormagnetogram is displayed in Figure~\ref{fig7} (top panel). Transeverse field vectors (arrows) are plotted on $B_z$ map. Note the sheared PIL exists only in the inner bipolar region enclosed by cyan rectangle region. Owing to strong shear in transeverse field vectors, the PIL generally spreads with intense vertical current ($J_z$) distribution as shown in the bottom panel.  In Figure~\ref{fig8}, at four representative epochs, we plot the velocity field derived from DAVE4VM on $B_z$ map. The FOV covers only inner bipolar region enclosed by a rectangle in Figure~\ref{fig7}, because it is the region of core field of the sigmoid being built before all CMEs. Vectors indicate the direction of flux motion within a region of polarity outlined by contour of $B_z$ ($\pm$130 G ). Different features move with different velocity at different epochs, where the velocity is spread upto a maximum value of 0.8 km/s consistent with previous studies of various tracking methods \citep{shibu2000,liy2004,liy2010, vemareddy2012a}. We normalised magnitude of velocity vectors to 0.5 km/s such that features that move at far less velocity will be recognised. As is obvious from this plot, the velocity field in negative polarity is coherent with a net organized flow pattern in first phase of evolution, which reflects a net southward motion of negative polarity with respect to positive polarity.
  
In the later instances this velocity pattern becomes converging motion towards positive polarity. We can see intruding negative polarity into positive polarity (panel at 18T17:00 UT) and its disappearance in successive panels. \citet{chae2002,chae2004} reported that submerging opposite polarity are the sites of cancelling magnetic features. In addition to these motions, a net northward motion in the top portion of positive polarity refers a similar effect of shearing flux motions. These predominant shearing and converging motion patterns upto June 23 are consistent with the net declining profiles of positive and negative flux in time. These observations evidence the persistent shearing and converging motions about the PIL played prime role in the cancellation of fluxes and a repeated formation of sigmoid in the corona, as studied in earlier sections \citep{martinsf1998,WangMuglach2007}. Especially, this AR observations are similar to those found in AR 8038 producing sequential CMEs in few days apart \citep{liy2010}.  

\begin{figure*}[!ht]
	\centering
	\includegraphics[width=.8\textwidth,clip=]{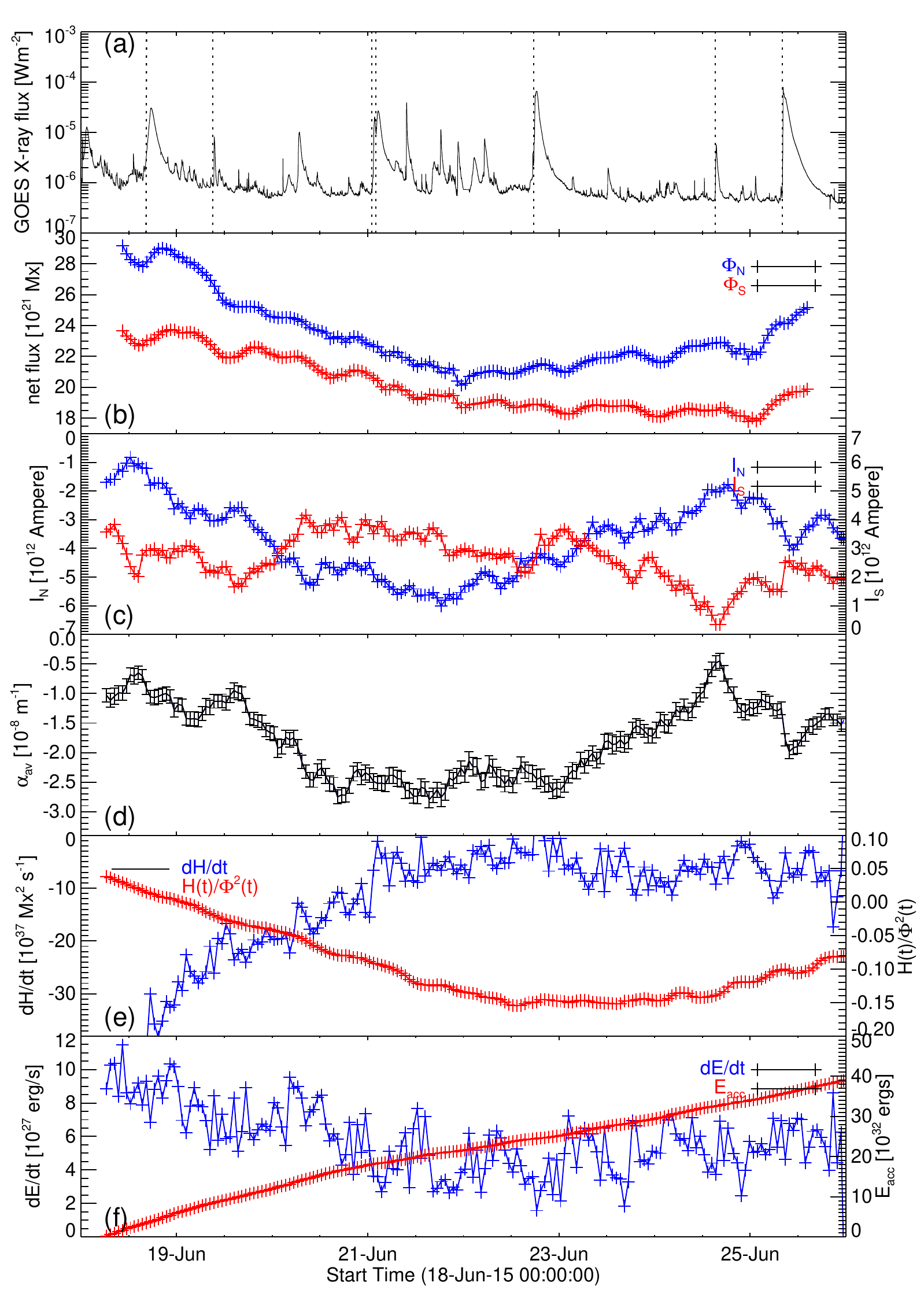}
	\caption{Time evolution of magnetic parameters in the AR 12371 during its disk transit a) disk integrated soft X-ray GOES flux, shows CME associated flares, b) net magnetic flux from north and south polarities, c) net vertical current from north and south polarities, d) twist parameter  $\alpha_{av}$, e) helicity flux injection, and coronal helicity $H(t)$ normalised by the mean magnetic flux, f) energy flux injection ($dE/dt$) and its coronal accumulation. Vertical dotted lines refer to start time of flares in this AR. Decrease in net magnetic flux represents flux cancellation about the PIL. Corresponding to the persistent strong shear motions, the H-flux is strong upto June 21 and accordingly $I$, $\alpha_{av}$ show increasing trend.}
	\label{fig9}
\end{figure*}

In Figure~\ref{fig9}, the time evolution of these parameters (18-25, June 2015) are plotted against GOES X-ray flux. The CME associated flares from this AR are prominent from the background (see Table~\ref{tab1}) and their initial timings are indicated by vertical dotted lines. Obviously, the profile of the net magnetic flux delineates a gradual decrease in both polarities during 18 to 23. This is consistent with the dispersing and disappearing polarities (Figure~\ref{fig1},\ref{fig8}). In the first five days of evolution, the positive (negative) flux decreased by $8(5)\times10^{21}$Mx, contributing to a 24\% net flux decrease. The rate of this decrease (cancellation) is significantly more than that in AR8038 (which is 18\%) in a period of 66hrs \citep{liy2010}. It also represents underlying faster and effective converging footpoint motions launching CMEs at lesser time interval than in AR 8038.

The net current ($I$) from both polarities and $\alpha_{av}$ show increasing trend till June 21 followed by stable evolution upto 23 June and then show decreasing nature till 25 June. This increased $I$ refers to increased stress in the magnetic field by shear/twist motions which accounts in horizontal field components and is a signature of AR eruptive behaviour \citep{falconer2002, schrijver2009, vemareddy2015c}. Note that the net current in north (south) polarity is negative (positive) (see Figure~\ref{fig7}), implying a dominant left hand twist to the field in accordance with the inverse-S sigmoid morphology of EUV observation as also interpreted in many previous studies (e.g., \citealt{vemareddy2012b, vemareddy2015a,vemareddy2017a}).  The last eruption occurred under a rapidly increasing conditions of net current, and $\alpha_{av}$ as a likely case due to the intermittent shear motions at a time scale of 4-8 hours. 

The profile of $dH/dt$ is having a value $-40\times10^{37}Mx^2s^{-1}$ in early observation period on June 18, which then decreases to an average value of $-5\times10^{37}Mx^2s^{-1}$ in the rest of the evolution time. Many studies found that the helicity flux dominantly comes from horizontal motions \citep{vemareddy2012a, vemareddy2015a} especially after the rapid emergence phase of the AR. As the AR is already emerged, therefore the observed profile of dH/dt indicates the magnitude of shear motions on the AR magnetic system is strong in the first phase than compared to later part of the evolution. This effect of strong shear motion in turn affects the $I$, $\alpha_{av}$ experiencing increased nature later. The effect of plasma motions on magnetic field observationally recognised only recently especially during unusual patterns of sunspot rotation, and generally the time evolution profile of $dH/dt$ correlates many non-potential parameters measured purely based on magnetic field \citep{vemareddy2012a,vemareddy2012b}. In this case, we see a time delay between fast injection of $dH/dt$ and the subsequent increase of $I$ and $\alpha_{av}$, indicating relaxation of magnetic field being driven after plasma motions \citep{vemareddy2017a}.  

\begin{figure*}[!ht]
	\centering
	\includegraphics[width=1.\textwidth,clip=]{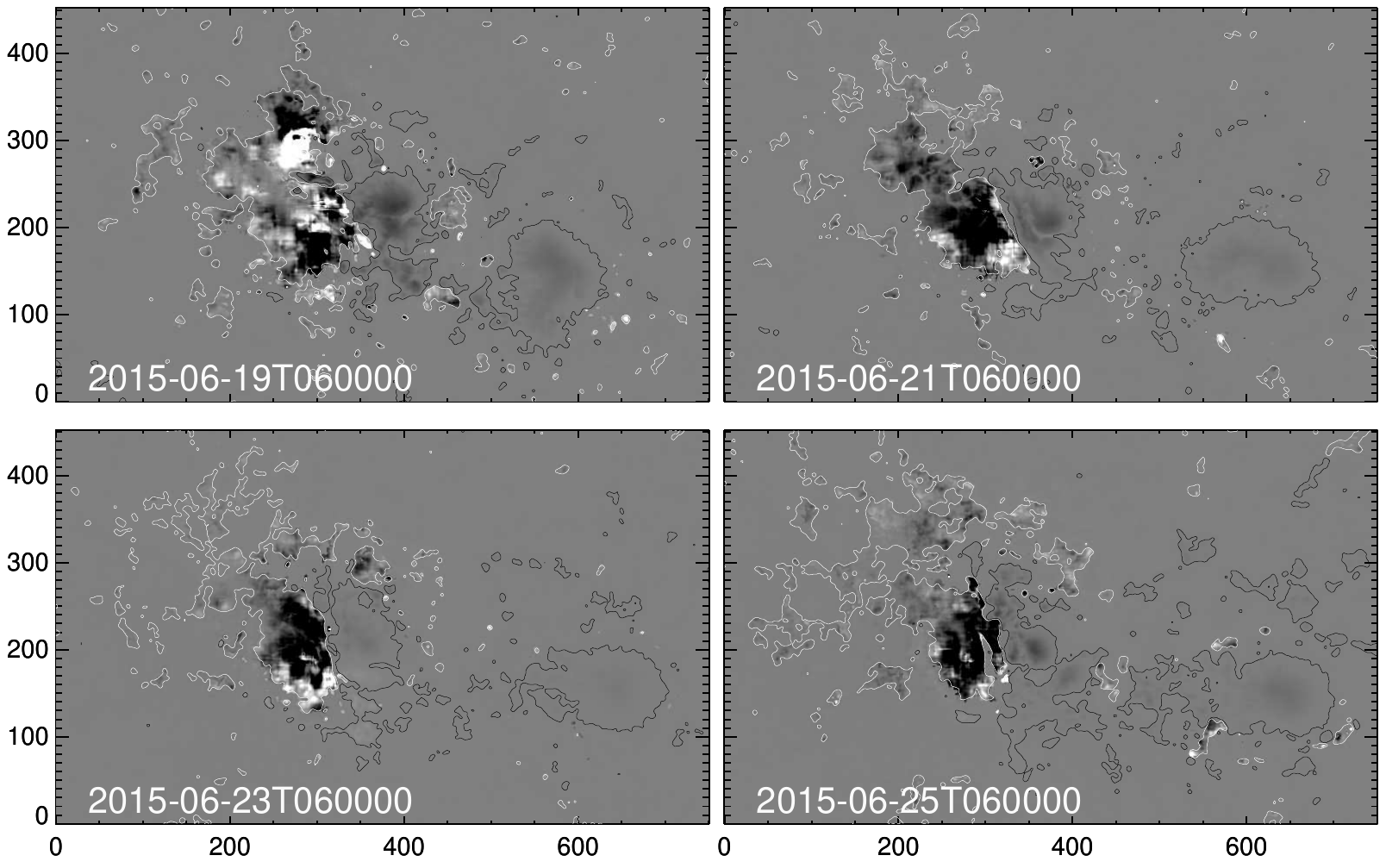}
	\caption{ Helicity flux distribution ($G_\theta$) across AR 12371 on four different days of evolution. Maps are scaled within $\pm10^{19}\,Mx^2cm^{-2}s^{-1}$. White (black) contours refer to 150(-150)G levels of $B_z$. Axis units are in pixels of 0.5arcsec. Intense signal of negative helicity flux distributed over positive polarity (P1, P2) due to shearing and converging motion of its negative counterpart.  }
	\label{fig10}
\end{figure*}

In Figure~\ref{fig10}, we plot the spatial distribution of $G_\theta$ ($dH/dt=\int{G_\theta(x,y)d\mathbf{S}}$), see for example \citealt{vemareddy2015a, vemareddy2017a}) at four different epochs of the AR evolution. White (black) patches refer positive (negative) sign of helicity flux and we scaled the maps within $\pm10^{19}Mx^2cm^{-2}s^{-1}$ so that all magnetic elements are  visible. These maps delineate intense negative H-flux distributed over the following positive polarity N2, N3, where sheared core field forms in due course of the AR evolution. By definition, the $dH/dt$ is the summation over all the photospheric elementary flux pairs of their net angular rotation around each other weighted by ${{B}_{n}}.B_{n}^{'}$ \citep{berger1988, pariat2005}. For example, if two positive (negative) end points rotate counter clockwise ($dH/dt > 0$) then their net contribution to $dH/dt$ is negative and consequently the field lines above them become twisted in a left-handed sense. Since the existing shear motion between opposite polarities about PIL refers to a net clock-clockwise rotation with respect to each other, the negative H-flux results at each spatial point due to weighting factor of opposite polarity product ${{B}_{n}}.B_{n}^{'}$. Here intense signal over positive polarity signifies a strong shear motion of negative polarity compared to that of positive polarity. The leading polarity shows only separation motion, so its H-flux distribution is feeble although negative in sign. Overall, the negative flux distribution is co-spatial with the sheared core of the sigmoid. Therefore, the observed $dH/dt$ in conjunction with AIA observations implies that the helicity flux being pumped by shear motions is utilised in the repeated buildup of sheared core sigmoid over the PIL, which then relaxes through intermittent explosions by formation of FR and its eruption. We also checked the $dH/dt$ profile calculated from DAVE \citep{schuck2005} method of tracked velocity field of LOS magnetic field observations, which yields an identical evolution trend as DAVE4VM (see also \citealt{demoulin2003,vemareddy2017a}). 

Coronal accumulation of helicity is obtained by time integration of helicity injection rate ($H=\int\limits_{0}^{t}{\frac{dH}{dt}\,}\Delta t$ ) over the observation time interval. In the plot, we show normalized helicity by average net magnetic flux ($\Phi$) over positive and negative polarities. $H/\Phi^2$ indicates how much the magnetic configuration is twisted/sheared, because for a uniformly twisted flux tube with $n$ turns, the helicity $H$ is equal to $n \Phi^2$, where $\Phi$ is its axial flux. From the time evolution, the AR flux system is twisted significantly upto 0.15 turns. Note that accumulated helicity is expelled through CMEs intermittently. Previous AR studies found a maximum twist of 0.2 turns \citep{demoulin2009, vemareddy2017a}  for highly twisted flux systems. Corresponding to $dH/dt$ profile, the energy flux injection $dE/dt$ (a positive definite quantity always) shows higher rate in the first phase (upto June 21) compared to later part of evolution where it varies about a mean value of $5\times10^{27}$ergs/s. Assuming an average time of 40 hours between two CMEs, this average energy injection is sufficient to store energy ($7.2\times10^{32}$ergs) required by a fast CME preceded by M-class flare. The coronal energy budget amounts to $3.9\times10^{33}$ergs, which is a result of purely horizontal flux motions during the considered time interval. This is as significant as to power the sequential CMEs with M-class flares and what matters here is how it availed to ejective eruptions rather slow confined eruptions dissipating as gradual coronal heating \citep{vemareddy2017a}.    

\section{Comparision with the magnetic evolution in non-eruptive AR 12192}
\label{comp}
\begin{figure*}[!ht]
	\centering
	\includegraphics[width=.75\textwidth,clip=]{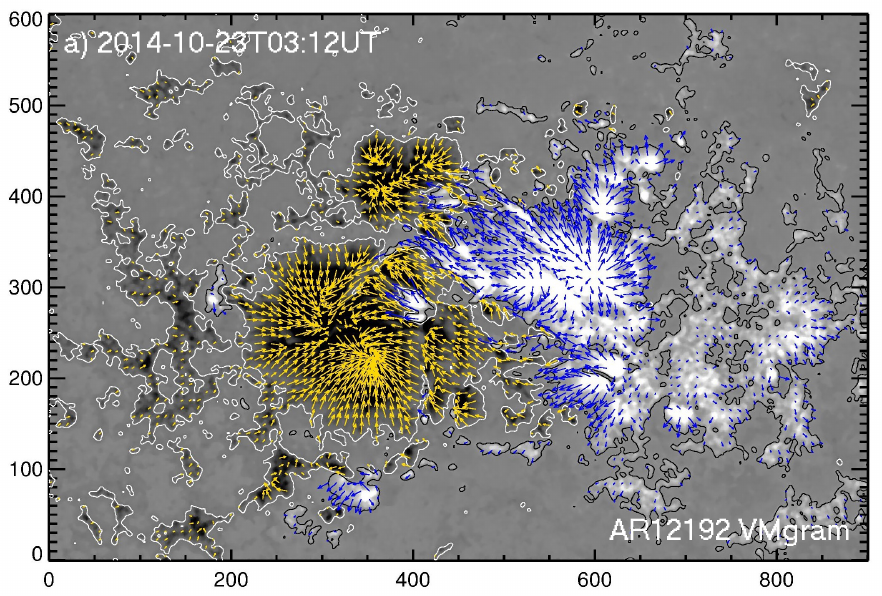}
	\includegraphics[width=.8\textwidth,clip=]{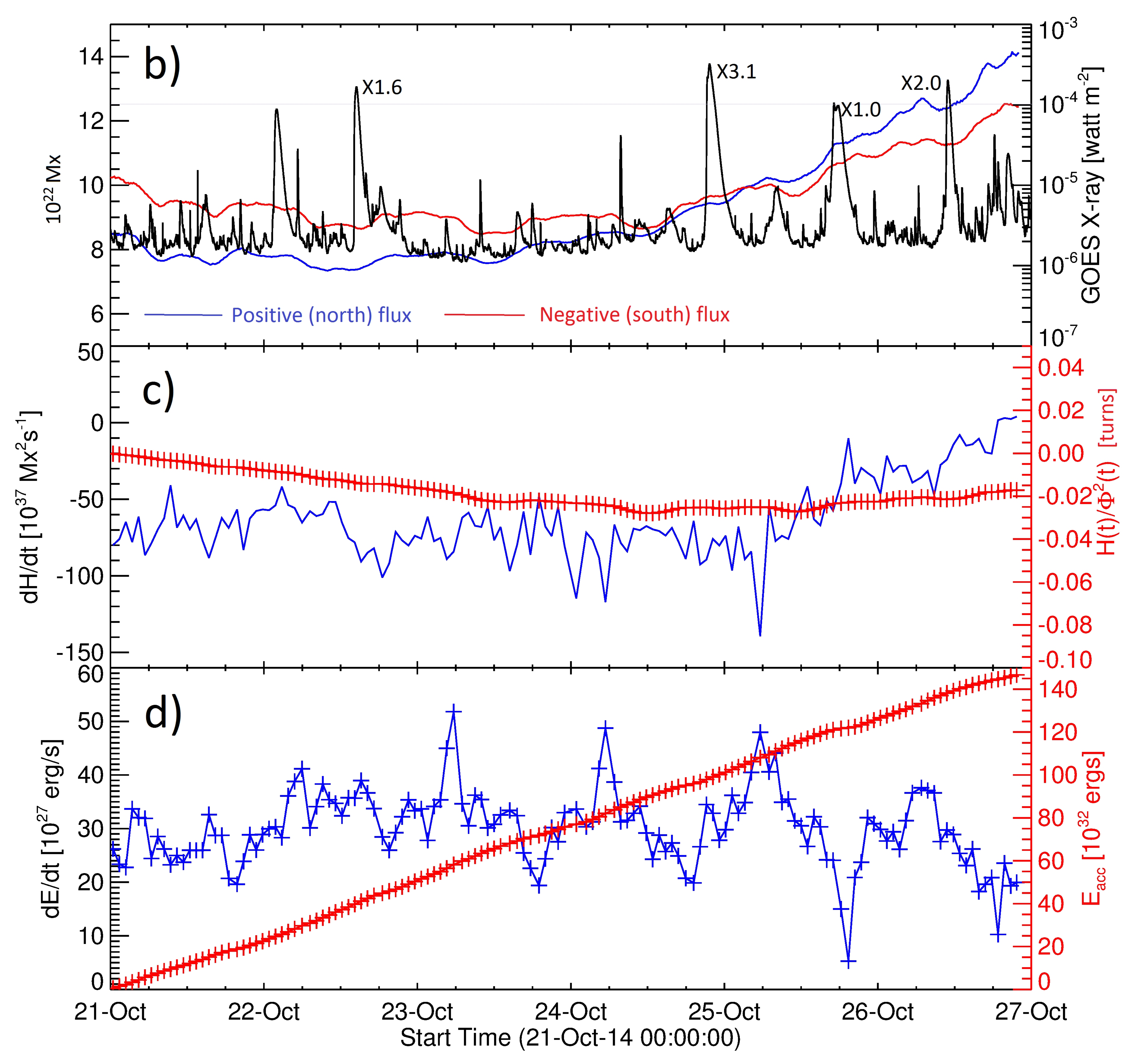}
	\caption{ Magnetic evolution in non-eruptive AR 12192. a) Vectormagnetogram showing field vectors on $B_z$ map,
		b) time evolution of net magnetic flux in south and north polarity. GOES-X ray flux is plotted with y-axis scale on right. X-flares from this AR are marked, c) time evolution of dH/dt and normalised coronal helicity. Note that $H(t)/\Phi^2$ variation is very low within -0.02 turns compared to AR 12371, d) evolution of $dE/dt$ and its accumulation. }
	\label{fig11}
\end{figure*}

Finally, we compared the magnetic evolution in AR 12371 with the largest AR 12192 in the past 24 years. The AR 12192 crossed the visible disk from 2014 October 17 to 30, unusually producing more than one hundred flares, including 32 M-class and 6 X-class ones, but only one small CME. Being flare rich but surprisingly CME poor, the AR 12192 had drawn much attention \citep{sunx2015,thalmann2015} in contrast to the AR case here. After applying similar methodology, the magnetic evolution of AR 12192 is plotted in Figure~\ref{fig11}. To be within $\pm45^\circ$ longitude, we consider the evolution for seven days between 21-27, during which four X-flares  occurred as shown in the top panel. The field distribution is largely bipolar with leading positive and following negative flux as shown in panel (a). Note that the magnetic evolution exhibits flux emergence from mid of June 24 with increased negative signed helicity flux distribution in the outer regions of main polarities. 

The $dH/dt$ varies about $-70\times10^{37}Mx^2s^{-1}$ till October 25, which is followed by persistent decreasing trend to zero value corresponding to the increased magnetic flux content in AR. In contrast to AR 12371 (Figure~\ref{fig9}), the $H(t)/\Phi^2$ is very small mounting to 0.02 turns (negative sign) in the seven day evolution. It is smaller by a factor 8 compared to that in AR 12371, although the net $dH/dt$ is twice in magnitude. It indicates that the AR 12192 has large average net flux content which is not part of twisted or shearing flux system. Note that $H(t)/\Phi^2$ can also be obtained by time integrating normalised $dH/dt$, which is a measure of helicity flux injection per unit flux unit time. It varies about the mean value of $0.1\times10^{-5}s^{-1}$ in AR 12192, $0.25\times10^{-5} s^{-1}$ in AR 12371. The energy flux $dE/dt$ changes about $30\times10^{27}$\,ergs/s, which is three times larger than that in AR 12371. However, the normalised $dE/dt$ by net average flux ($\Phi$) are comparable, $0.25\times10^{6}\,ergs/Mx/s$ (average over time), $0.3\times10^6\,ergs\,Mx^{-1} s^{-1}$ in AR 12371, AR 12192 respectively. In total, the net $dH/dt$, $dE/dt$ are larger in AR 12192 due to its supersize but the normalised coronal helicity infers a weakly twisted flux which seems to be the prime indicator for being CME poor. This inference is consistent with the results of \citet{sunx2015} suggesting the weaker non-potentiality and stronger overlying field in AR 12192.  
 
\section{Discussion and Conclusion}
\label{disc}
All four CMEs occurred from the same magnetic PIL during disk passage of the AR12371. Under the same evolving conditions of magnetic field, the sheared core about the PIL (sigmoid) forms repeatedly in several hours apart. The observations of coronal morphology implies internal tether cutting reconnection as the suggestive triggering mechanism during all four CMEs, suitably explaining the pre-cursor brightening, formation of FR and the eventual eruption. All associated flares are M-class LDEs ($>4$hrs) transforming sheared core sigmoid to less sheared arcade. The EUV double (symmetric) dimming, three part CME structure, and fast propagation speeds in LASCO FOV altogether characterize the CMEs as homologous events.    

Flux motions parallel to PIL in opposite direction are very primitive to build stress in straddled potential field arcade \citep{antiochos1994}. The field lines near the PIL become parallel to PIL (sheared core) and those away from PIL are less sheared envelope. In such a magnetic structure, converging motions ignite a slow reconnection of field lines at the opposite ends of PIL producing helical field lines \citep{pneuman1983,ballegooijen1989}, which in turn manifest a FR. This is a process associated with flux cancellation and occurs over several hours to even days. The dips of helical fields accommodate the chromospheric material lifting against gravity and explains the observed dark filament channels in EUV 304\AA~images, for example in CME2 case (Figure~\ref{fig4}). In the FR models, weakly twisted FRs are considered as sheared arcades, in the sense that the magnetic field is dominated by the axial component \citep{mackay2010}, which is the case in AR 12371 before all observed eruptions.

In case highly twisted FR forms from the sheared arcade configuration before the eruption by the above slow reconnection, there can arise other instabilities like ideal-kink instability and/or torus instability which play the role of CME initiation. In fact, all FR models assume an analogous AR magnetic configuration with pre-formed FR and observationally recognizing a clear FR topology is crucial to disentangle the triggering mechanism whether reconnection or ideal MHD instability related. Observational studies \citep{green2007, lynch2009, vemareddy2016b} suggest the FR formation/augmentation from sheared core sigmoid under the evolving conditions of magnetic flux cancellation. They also reported the possibility of tether-cutting reconnection as a triggering mechanism of CME eruption. In our AR12371,  the post flare arcade from previous eruption becomes strong sheared core sigmoid with two opposite J-sections by shearing and converging motions and the FR forms only during the onset of eruption by a self-amplifying internal tether-cutting reconnection.

The entire magnetic evolution of the AR is better explained in terms of magnetic energy storage and release process. Coronal X-ray sigmoid is a stored energy configuration and the most recognized eruptive feature \citep{canfield1999, moore2001,green2009}. During the eruption, the sigmoid transforms to a pattern termed ``sigmoid-to-arcade” indicating coronal structure changes from a highly sheared to less sheared (or PFA) magnetic configuration. This reconfiguration releases the magnetic energy that is being pre-stored in stressed form by persistent shearing motions as the case here. After this reconfiguration, a restore process begins to convert the PFA again to sheared core sigmoid the so-called ``arcade-to-sigmioid” evolution. Depending on the nature of flux motions, the time scale of energy storage varies. In AR12371, it is 31(CME1 \& CME2), 40(CME2 \& CME3), 60hrs (CME3 \& CME4) respectively between successive CMEs. This process of ``sigmoid to arcade to sigmoid” continues in a cylce for the repeated formation of sigmoid and its eruption, observing as successive CMEs like this AR. Observational study by \citet{liy2010} aslo indicated this cyclic process in action causing sequential CMEs from AR 8038.

The AR evolution including observed CME eruptions, can also be explained in terms of magnetic helicity. In our case, the AR coronal magnetic field is mainly driven by continued shear motions, which inject helicity flux of dominant negative sign (Figure~\ref{fig9}). This sheds into sigmoid’s sheared core twist and its energy by building a twisted FR. Note that FR can also form during the onset of eruption by tether-cutting reconnection. In the case of pre-formed FR, ideal MHD instability can trigger the eruption. In either of the cases the sheared core sigmoid, being the precursor structure in all four studied cases  in AR 12371, is basic structure with a critical amount of volume helicity. As the coronal field can not accommodate indefinite amount of helicity, the only way to get rid of it is to expel into interplanetary space by a CMEs  \citep{low1994,zhangmei2005, zhangmei2013}. It occurs by FR bodily ejection initiated by the above said instability. On the other hand, if the helicity injection is changing sign over the period of AR evolution, pre-accumulated helicity of one sign is cancelled by the opposite sign helicity flux in later time. This cancellation of coronal helicity is manifested by gradual field reconfiguration and dissipation of energy heating the corona (no FR). Using HMI vector magnetic field observations of emerging ARs, \citet{vemareddy2015a} reported these two possibilities of coronal magnetic field evolution. He showed three kinds of AR evolution with a net positive, negative, and successive injection of positive and negative helicity flux. The ARs with a predominant sign launch CMEs at some point of time. However, the AR with successive injection of opposite helicity exhibits only C-class flaring activity characterised by delayed enhanced coronal emission with respect to the time of sign change of helicity flux \citep{vemareddy2017a}. However, it is yet to see this theoretical relation of coronal helicity flux and the eruptive phenomena in several emerging AR cases.      

Notably, the predominant sign of helicity flux over time seems not a only requirement for an AR to be CME productive. As shown in Figure~\ref{fig11}, the AR 12192 is having same sign with large injection of helicity flux over time but CME poor as a counter case to our AR 12371. The $dE/dt$ {\bf is} comparable in both of these confined and eruptive ARs.  Even being predominant sign injection, a major difference seen is with $H(t)/\Phi^2$ indicating weakly twisted flux in AR 12192 and which seem to be the indicator of eruptiveness of an AR. Note that $H(t)/\Phi^2$ is time integrated quatity of $\frac{1}{\Phi^2}\frac{dH}{dt}$ which is helicity flux injection per unit magnetic flux per unit time. A weakly twisted flux in an AR does not represent a flux rope and it is unlikely that the magnetic evolution in AR 12192 is in favor of flux rope formation. This is in agreement with the study by \citet{jiangc2016b} indicating the AR remained in shared arcade configuration without forming two-J shape like and escaping flux rope unlike in many sigmoid ARs, including the one here. Therefore, we propose to use normalised helicity flux as a measure for AR eruptiveness accommodating the role of background flux. \citet{sunx2015} points that weak non-potentiality and strong background field as the reasons for confined nature of AR 12192. In moderate size ARs, small value of $H(t)/\Phi^2$ implies to large flux content unrelated to sheared/twisted part which can act as overlying flux. Since an eruption occurs either due to weak overlying field or strong flux rope, higher value of $H(t)/\Phi^2$ likely forms a flux rope with weak overlying flux. In short, a small (larger) value of $H(t)/\Phi^2$ implies weakly (strongly) twisted AR flux system likely containining dominant overlying (flux rope related) flux suppressing (favoring) the eruption.  However, the upper limit of $H(t)/\Phi^2$ is yet to be evaluated from a study of statically significant AR cases over which the CMEs are inevitable.         

The AR 12371 is a representative for eruptive class ARs as its couterpart AR 12192 with no CMEs. While highlighting the favored trigger mechanism under a given boundary flux motions, our study of magnetic evolution provides clues for differentiating eruptive and non-eruptive ARs in terms of magnetic helicity injection. To further substantiate the observational results and interpretations, we intend to study the modeling aspect of the AR magnetic structure and the related topological aspects.

\acknowledgements SDO is a mission of NASA's Living With a Star Program. SOHO is a project of international cooperation between ESA and NASA. This work used the DAVE4VM code written and developed by P. W. Schuck at the Naval Research Laboratory. I thank the referee, Prof. P. D\'emoulin, Prof. B. Ravindra for useful comments and suggestions.  P.V. is supported by an INSPIRE grant of AORC scheme under the Department of Science and Technology.  

\bibliographystyle{apj}

\end{document}